\documentclass[
]{ceurart}

\usepackage{listings}
\begin{document}

\copyrightyear{2024}
\copyrightclause{Copyright for this paper by its authors.
  Use permitted under Creative Commons License Attribution 4.0
  International (CC BY 4.0).}

\conference{PhysioCHI: Towards Best Practices for Integrating Physiological Signals in HCI,   May 11, 2024, Honolulu, HI, USA}

\title{Exploring the Optimal Time Window for Predicting Cognitive Load Using Physiological Sensor Data}

% \tnotetext[1]{You can use this document as the template for preparing your
%   publication. We recommend using the latest version of the ceurart style.}


\author{Minghao Cai}[%
orcid=0000-0002-3331-5979,
email=minghaocai@ualberta.ca,
]

\author{Carrie {Demmans Epp}}[%
orcid=0000-0001-9079-4921,
email=cdemmansepp@ualberta.ca,
]

\address{EdTeKLA Research Group, Department of Computing Science, University of Alberta, Edmonton, Canada}

% \author[1]{Dmitry S. Kulyabov}[%
% orcid=0000-0002-0877-7063,
% email=kulyabov-ds@rudn.ru,
% url=https://yamadharma.github.io/,
% ]
% \cormark[1]
% \address[1]{Peoples' Friendship University of Russia (RUDN University),
%   6 Miklukho-Maklaya St, Moscow, 117198, Russian Federation}
% \address[2]{Joint Institute for Nuclear Research,
%   6 Joliot-Curie, Dubna, Moscow region, 141980, Russian Federation}

% \author[3]{Ilaria Tiddi}[%
% orcid=0000-0001-7116-9338,
% email=i.tiddi@vu.nl,
% url=https://kmitd.github.io/ilaria/,
% \address[3]{Vrije Universiteit Amsterdam, De Boelelaan 1105, 1081 HV Amsterdam, The Netherlands}

% \author[4]{Manfred Jeusfeld}[%
% orcid=0000-0002-9421-8566,
% email=Manfred.Jeusfeld@acm.org,
% url=http://conceptbase.sourceforge.net/mjf/,
% \address[4]{University of Skövde, Högskolevägen 1, 541 28 Skövde, Sweden}

% %% Footnotes
% \cortext[1]{Corresponding author.}
% \fntext[1]{These authors contributed equally.}

\begin{abstract}
Learning analytics has begun to use physiological signals because these have been linked with learners' cognitive and affective states. These signals, when interpreted through machine learning techniques, offer a nuanced understanding of the temporal dynamics of student learning experiences and processes. However, there is a lack of clear guidance on the optimal time window to use for analyzing physiological signals within predictive models. We conducted an empirical investigation of different time windows (ranging from 60 to 210 seconds) when analysing multichannel physiological sensor data for predicting cognitive load. Our results demonstrate a preference for longer time windows, with optimal window length typically exceeding 90 seconds. 
These findings challenge the conventional focus on immediate physiological responses, suggesting that a broader temporal scope could provide a more comprehensive understanding of cognitive processes. In addition, the variation in which time windows best supported prediction across classifiers underscores the complexity of integrating physiological measures. Our findings provide new insights for developing educational technologies that more accurately reflect and respond to the dynamic nature of learner cognitive load in complex learning environments.
\end{abstract}

\begin{keywords}
  Cognitive Load\sep
  Physiological Signals \sep
  Learning Analytics \sep
  User Modelling
\end{keywords}

\maketitle

\section{Introduction}

Physiological signals have long been used within learning analytics research because these signals are associated with learners' underlying cognitive and affective states. Recent advances in machine learning and sensors have enabled researchers to adopt a more nuanced approach that accounts for temporality. This more nuanced approach supports improved understanding of underlying, latent constructs that are important to learning. These constructs include cognitive load and its impact on performance \cite{dmello_advanced_2017,baker_impact_2010}.
However, there is a lack of established evidence for the choice of time window when using physiological signals in predictive modeling. In previous research, there has been a focus on immediate physiological responses to discrete tasks or events, often employing a narrow time window (e.g., 2s) to isolate specific responses directly attributable to such activities \cite{krejtz_eye_2018}. While this provides insight into a learner's immediate cognitive reactions, this approach may not fully reflect the complexities of learning processes that unfold over extended periods, especially in more interactive and dynamic settings \cite{sweller_development_2023}.

We investigated how varying the amount of sensor signal (time window length) affects model performance when using multichannel physiological sensors to predict cognitive load in complex learning environments. We conducted a empirical study assessing the impact of different window lengths when performing prediction with different classifiers. 
By extending the observation period, we aim to capture immediate physiological reactions as well as the gradual changes and patterns that emerge over time because these changes characterize learning processes \cite{xie_review_2000}.

\section{Related Work}

Cognitive Load Theory (CLT) is grounded in research about working memory and mental effort \cite{salomon_television_1984}. CLT provides a framework for understanding the link between cognitive demands and learning performance \cite{sweller_cognitive_2011, paas_cognitive_2003, paas_cognitive_2014}. It consists of three types of load: intrinsic, extraneous, and germane. Intrinsic load relates to the difficulty of the subject matter and varies with the learner's expertise. Extraneous load hinders learning and is a consequence of system or activity designs that cause unnecessary processing. Germane load supports learning via schema development and is under scrutiny because it is difficult to distinguish from the other subtypes of cognitive load. 

Cognitive load measurement has primarily used self-reports and task-based assessments. However, self-reports miss unconscious cognitive processes and are prone to recall bias, affecting the accuracy of reported cognitive effort \cite{krell_editorial_2022,leppink_development_2013}. Task-based methods are also limited. They evaluate cognitive load through reaction times or learner accuracy on secondary tasks that learners are asked to perform while completing the primary activity.
%, like responding to a light while engaged in another task. 
These task-based methods may impact performance on the primary task \cite{brunken_direct_2003} even though they are intended to use few cognitive resources.

Recently, the use of physiological methods, particularly pupil diameter, has been gaining attention. Pupil dilation is associated with increased cognitive challenges \cite{beatty_task-evoked_1982} and can be measured via eye-tracking technologies. This method involves comparing pupil size to a baseline, which can be affected by factors like lighting and camera angle \cite{beatty_pupillary_2000}. To overcome these issues, researchers developed the Index of Cognitive Activity (ICA) \cite{marshall_index_2002} and the Index of Pupillary Activity (IPA)  \cite{duchowski_index_2018}, focusing on real-time pupil fluctuations to provide more accurate assessments of cognitive load. These innovations offer improved reliability over traditional baseline comparisons by accounting for changes that may occur. 

Despite advances in measuring cognitive load through physiology, there remains a gap in the literature regarding the time window that should be employed when using physiological signals in predictive modeling. Prior studies have predominantly concentrated on capturing short-term physiological responses to distinct tasks or events, often employing narrow time windows. This approach helps link the measured responses to the cognitive tasks being examined. For example, Chen and Epps used a 12-second window for measuring the average size of the pupil to estimate cognitive load \cite{chen_using_2014}. 
In contrast, some have adopted a coarse, time-aggregated approach that consolidates the average pupil size from prolonged tasks into a single metric, typically overlooking nuances such as the timing, magnitude, or form of specific responses \cite{klingner_measuring_2008}. 
This approach introduces questions regarding the precision and efficacy of these measurements, as extended observation periods might blend specific cognitive load indicators with unrelated data.

Not surprisingly, we have yet to form a consensus on how to best select or adjust the time window to ensure models accurately reflect cognitive load. The selection of the time window is critical, as it can influence the reliability and validity of model inferences. A time window that is too brief may not fully reflect the extent of the cognitive load experienced during learning, whereas one that is overly extended could include irrelevant information, thereby complicating the task of pinpointing the exact cognitive demands placed on individuals. The absence of established guidelines for which window size to use necessitates further research. So, we explored the tuning of the time window for physiological signals and tested their differential predictive performance. 
In this paper, three channels of sensor data (pupil diameter, electrodermal activity [EDA], and heart rate) were used to predict cognitive load. 

\section{Methods}

\subsection{Dataset}
The dataset used in this study was collected from 35 English learners who interacted with an online literacy game for one hour \cite{cai_toward_2024}. Due to significant sensor noise, data from one participant was excluded, resulting in a final sample of 34 individuals aged 17 to 33 years ($M = 24.1$ years). None of the participants had prior experience with the game.

The dataset had two parts: physiological sensor data and self-reported cognitive load.

Physiological data were collected using a two-sensor system: an open-source eye tracker (Pupil Core, Pupil Labs) and a wireless wristband (E4, Empatica Inc.). The eye tracker, which has the form factor of standard glasses, captured eye movements and pupil dynamics at 200 Hz, alongside a scene camera recording the user's viewpoint. The wristband, worn like a watch, collected EDA data at 4 Hz. The wristband also collected heart rate from blood volume pulse (BVP) readings. Both BVP and EDA were measured in 10-second intervals. This setup allowed for the synchronous recording of learners' physiological responses during gameplay.

Self-reported cognitive load was measured approximately every 5 minutes. Scores ranged from 1 to 10 ($M = 5.0$, $SD = 1.83$) and were categorized into three levels for classification: low (1 to 3.33), moderate (3.34 to 6.66), and high (6.67 to 10). These levels were used as labels for the predictive model.

\subsection{Pre-processing of Physiological Data }
Pupil diameter signals for both eyes along with blink activity records were exported. This data included time-stamped pupil diameters and an associated confidence level, ranging from 0.0 (no detection) to 1.0 (high certainty). To prepare the data, we removed segments corresponding to blinks and instances of low confidence (below 0.65).
For data continuity, missing data due to blink and low-confidence removal were linearly interpolated. To further reduce noise, we applied a third-order Butterworth filter with a 4 Hz cutoff to minimize high-frequency disturbances while retaining signal fidelity \cite{krejtz_eye_2018}.

The EDA and heart rate signals were smoothed using a Simple Moving Average (SMA) filter. This  technique helps to minimize high-frequency noise while preserving the characteristics of the signals that are relevant for assessing arousal \cite{posada-quintero_innovations_2020}.

Post-cleaning, we segmented the physiological data (pupil diameter, EDA, heart rate), aligning it with the timestamps from participants' self-reported cognitive load data. We segmented the sensor data around these timestamps to reflect distinct periods of system interaction. 

Data recorded during self-report completion was excluded to avoid confounding effects from cognitive processes related to data collection. This process resulted in 312 segments of data.

\subsection{Experimental Procedures}

We experimented with different time windows for extracting physiological features. Features were extracted for each data segment. We tested the predictive performance of physiological data from 6 windows: 60s, 90s, 120s, 150s, 180s, and 210s.

For each time window, we extracted 7 features from the pupil diameter data: 
the frequency of changes in pupil diameter (PCF);
the minimum pupil size in the dataset (MinPD);
the average size of the pupil diameter (AvgPD);
the maximum pupil size in the dataset  (MaxPD);
the average speed at which the pupil diameter changes (AvgPV);
the maximum speed at which the pupil diameter changes (MaxPV); and
the largest continuous change in pupil diameter without direction change (MaxPC).

For each time window, we extracted 9 features from the EDA data: 
the frequency of changes in EDA (ECF);
the maximum continuous change in EDA without direction change (MaxEC);
the minimum value of the EDA (MinE);
the maximum value of the EDA (MaxE);
the average value of the EDA (AvgE);
the standard deviation of the EDA (SDGE);
the average speed of change in EDA levels over time (AvgEV);
the maximum speed at which the EDA changes (MaxEV);
the difference between the maximum and minimum EDA (RngE).

For each time window, we extracted 9 features from the heart rate data: 
the frequency of changes in heart rate (HCF);
the minimum value of the heart rate (MinH);
the maximum value of the heart rate (MaxH);
the average value of the heart rate (AvgH);
the standard deviation of the heart rate (SDH);
the average speed of change in heart rate (AvgHV);
the maximum speed at which the heart rate changes (MaxHV);
the difference between the maximum and minimum heart rate (RngH);
the maximum continuous change in heart rate without direction change (MaxHC).

We tested the predictive capabilities of each window with six machine learning classifiers: Naive Bayes (NB), Decision Tree (DT), Linear Support Vector Machine (Linear SVM), Radial Basis Function Support Vector Machine (RBF SVM), Logistic Regression (LR), and Random Forest (RF). 

The classification was subject-independent. We allocated 80\% of data for training and 20\% for testing. During training, we performed hyperparameter tuning with 4-fold cross-validation. Grid search was used to find the optimal settings, with tuning focusing on maximizing Cohen's $\kappa$ \cite{cohen_coefficient_1960}.

\section{Results and Discussion}

\begin{table}[h]
\centering
\caption{Model Performance and window size}
\begin{tabular}{llccc}
\hline
\multicolumn{2}{l}{\textbf{Classifier}} & \textbf{$\kappa$} & \textbf{Accuracy} & \textbf{Window (s)} \\
\hline

& Naive Bayes & .107 & .417 & 90 \\
& Decision Tree & .211 & .600 & 150 \\
& Linear SVM & .218 & .533 & 120 \\
& RBF SVM & .279 & .617 & 150 \\
& Logistic Regression & .276 & .534 & 210 \\
& \textbf{Random Forest} & \textbf{.434} & \textbf{.700} & \textbf{210} \\
\hline
\multicolumn{5}{l}{\footnotesize{Bold font indicates the best result}} \\
\end{tabular}
\label{table:combined_model_performance_window}
\end{table}

\begin{figure}[]
    \centering
    \includegraphics[width=0.75\linewidth]{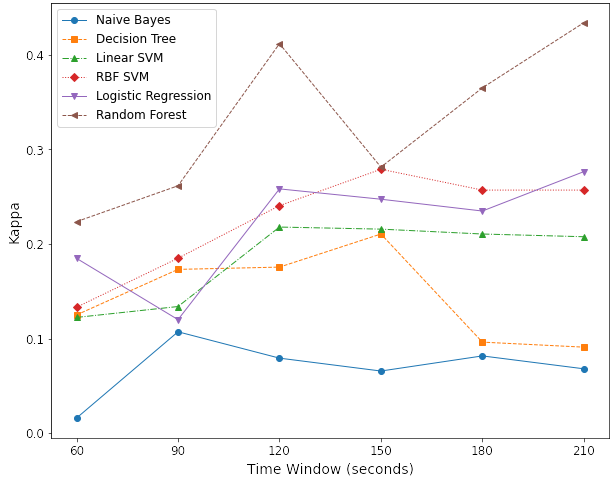}
    \caption{Kappa values for each model by window size.}
    \label{fig:kappas}
\end{figure}

Table \ref{table:combined_model_performance_window} reports the time window where the model performed best. Most models performed best when the window was longer than 90s and the best performing model (Random Forest) used a window that exceeded 3 minutes. 

When we inspected model performance across windows, it became clear that window size impacted classifier performance (Figure \ref{fig:kappas}). Our results indicated that peak performance for all classifiers was achieved when 90 seconds or more of sensor data were used as input to the model.
Most classifiers, apart from Naive Bayes, demonstrated improved performance with wider windows.

Our findings suggest that when it comes to predicting cognitive load in interactive game-based learning settings, classifiers tested with a wider window (150 or 210) generally performed better. This outcome challenges previous research that prioritized immediate physiological reactions using shorter time-frames. Learning sessions, especially in complex environments like games, can last longer. Our findings suggest that a larger time window may allow comprehensive tracking of physiological changes, potentially reflecting the gradual increase and decrease of cognitive load as learners navigate through different interactive events. Such settings might include the impact of delayed physiological responses or the accumulation of mental fatigue that shorter windows may not capture. This shift in methodology not only enriches our understanding of cognitive processes in complex learning environments but also enhances the potential for developing more effective educational technologies that are sensitive to the fluctuating dynamics of learner engagement and cognitive load. 

Furthermore, the variation in optimal time windows among different classifiers highlights the complexity of integrating physiological measures for cognitive load assessment. This variability suggests that the best analysis window differs based on several factors, including the educational setting, task complexity, types of physiological data, and the specific algorithms used. Therefore, a universal approach to time window selection for all scenarios seems impractical. Instead, customizing the time window based on the algorithm, context, and learning environment could yield more accurate and meaningful insights into cognitive load dynamics during educational activities.

\section{Conclusion}

Learner physiology is an important resource for understanding changes in internal states, such as cognitive load. Yet, there exists a lack of evidence that can inform time window selection when using physiological signals in predictive modeling, particularly within complex educational environments. In this study, we conducted an experiment to explore model performance when using different window sizes to predict cognitive load from physiological data. Our study provides insight into how to employ physiological signals in educational technology research by demonstrating that a wider time window is beneficial for modelling some constructs, i.e., cognitive load. We posit this benefit is due to the extended sensor data's ability to capture cognitive adaptation. Most importantly, the presented analyses highlight the potential interactions between algorithms and the amount of physiological sensor data used.  By understanding how to select an appropriate time window for physiological data, researchers and practitioners can optimize the data collection process thereby enhancing both the precision and reliability of the systems used to monitor and respond to student cognitive states. It opens up possibilities for creating more personalized learning experiences that adjust content difficulty and presentation using real-time assessments of cognitive load.

\begin{acknowledgments}
This work was supported in part by funding from the Social Sciences and Humanities Research Council of Canada and the Natural Sciences and Engineering Research Council of Canada (NSERC), [RGPIN-2018-03834].

\end{acknowledgments}

%%
%% Define the bibliography file to be used
% \bibliography{references}

% \appendix


% The sources for the ceur-art style are available via
% \begin{itemize}
% \item \href{https://github.com/yamadharma/ceurart}{GitHub},
% % \item \href{https://www.overleaf.com/project/5e76702c4acae70001d3bc87}{Overleaf},
% \item
%   \href{https://www.overleaf.com/latex/templates/template-for-submissions-to-ceur-workshop-proceedings-ceur-ws-dot-org/pkfscdkgkhcq}{Overleaf
%     template}.
% \end{itemize}

\end{document}